\documentclass{elsart}

\usepackage{graphicx}
\usepackage{eucal}
\usepackage[dvips]{epsfig}
\usepackage{amssymb}
\usepackage{amsmath}
\usepackage{cite}

\begin{document}



\newcommand{\be}{\begin{eqnarray}}
\newcommand{\ee}{\end{eqnarray}}
\newcommand{\bse}{\begin{subequations}}
\newcommand{\ese}{\end{subequations}}


\newcommand{\bnum}{\begin{enumerate}}
\newcommand{\enum}{\end{enumerate}}

\newcommand{\bit}{\begin{itemize}}
\newcommand{\eit}{\end{itemize}}

\newcommand{\bc}{\begin{cases}}
\newcommand{\ec}{\end{cases}}


\newcommand{\erf}{\mrm{erf}}


\newcommand{\bpm}{\begin{pmatrix}}
\newcommand{\epm}{\end{pmatrix}}

\newcommand{\bvm}{\begin{vmatrix}}
\newcommand{\evm}{\end{vmatrix}}


\newcommand{\bs}{\boldsymbol}
\newcommand{\mbb}{\mathbb}
\newcommand{\mcal}{\mathcal}
\newcommand{\mfr}{\mathfrak}
\newcommand{\mrm}{\mathrm}

\newcommand{\ul}{\underline}
\newcommand{\ovl}{\overline}



\newcommand{\ga}{\alpha}
\newcommand{\gb}{\beta}
\newcommand{\gc}{\gamma}
\newcommand{\gd}{\delta}
\newcommand{\eps}{\epsilon}
\newcommand{\gf}{\phi}
\newcommand{\gl}{\lambda}
\newcommand{\gk}{\kappa}
\newcommand{\go}{\omega}
\newcommand{\gt}{\theta}
\newcommand{\gr}{\rho}

\newcommand{\Gf}{\Phi}
\newcommand{\Go}{\Omega}
\newcommand{\Gc}{\Gamma}
\newcommand{\Gt}{\Theta}
\newcommand{\Gd}{\Delta}

\newcommand{\gve}{\varepsilon}
\newcommand{\gvf}{\varphi}
\newcommand{\gvr}{\varrho}

\newcommand{\h}{\hbar}

\newcommand{\p}{\partial}
\newcommand{\f}{\frac}
\newcommand{\diff}{\mrm{d}}
\newcommand{\iy}{\infty}
\newcommand{\lap}{\triangle}
\newcommand{\nab}{\nabla}

\newcommand{\parder}[2]{\frac{\partial #1}{\partial #2}}
\newcommand{\totder}[2]{\frac{\mrm{d} #1}{\mrm{d} #2}}

\newcommand{\R}{\mathbb{R}}
\newcommand{\N}{\mathbb{N}}
\newcommand{\Z}{\mathbb{Z}}

\newcommand{\T}{\bs{\mathcal{T}}}

\newcommand{\fa}{\forall\;}
\newcommand{\ex}{\exists\;}
\newcommand{\lan}{\langle}
\newcommand{\ran}{\rangle}


\newcommand{\csp}{\;,\qquad}


\begin{frontmatter}


\title{Phase Transitions in Small Systems: Microcanonical
vs. Canonical Ensembles}
\author{J\"orn Dunkel}
\address{Institute for Physics, Universit\"at Augsburg, Universit\"atsstra{\ss}e 1,  D-86135 Augsburg, Germany}
\ead{joern.dunkel@physik.uni-augsburg.de}
\ead[url]{www.physik.uni-augsburg.de/$\sim$dunkeljo}
\author{Stefan Hilbert}
\address{Institute for Physics, 
Humboldt-Universit\"at zu Berlin, 
Newton-Stra{\ss}e 15,
D-12489 Berlin,
Germany}


\begin{abstract}
We compare phase transition(-like) phenomena 
in small model systems for both microcanonical and 
canonical ensembles. The model systems correspond to a few
classical (non-quantum) point particles confined in a one-dimensional
box and interacting via Lennard-Jones-type pair potentials. By means
of these simple examples it can be shown already that the microcanonical
thermodynamic functions of a small system may exhibit rich oscillatory
behavior and, in particular, singularities (non-analyticities) separating different
microscopic phases.  These microscopic phases may be identified as
different microphysical  dissociation states of the small system. 
The microscopic oscillations of microcanonical thermodynamic quantities
(e.g. temperature, heat capacity, or pressure) should in principle be 
observable in suitably designed evaporation/dissociation
experiments (which must realize the physical preconditions
of the microcanonical ensemble). By contrast, singular phase
transitions cannot occur, if a small system is embedded into an infinite heat
bath (thermostat), corresponding to the canonical ensemble. For the
simple model systems under consideration, it is nevertheless possible to
identify a smooth canonical phase transition by studying the  
distribution of complex zeros of the canonical partition function.    
\end{abstract}

\begin{keyword}
microscopic phase transitions\sep 
small systems\sep
Lennard-Jones chains 

\PACS 
05.70.Ce\sep 
05.70.Fh\sep 
05.70.Jk\sep  
64.60.Cn  
\end{keyword}
\end{frontmatter}


\section{Introduction}
\label{s:intro}

One of the most intriguing thermodynamic properties of various
macroscopic systems is their ability to undergo phase transitions
(PTs) if one or more control parameters pass certain critical
values \cite{Ka67,Le99}. The first systematic classification scheme
for macroscopic PTs was proposed by P. and T. Ehrenfest\cite{Eh12} in
1912 already. After further pioneering work by Mayer et
al. \cite{Ma37,MaAc37,MaHa38,StMa39}, Yang
and Lee \cite{YaLe52,LeYa52} elucidated the
mathematical essence underlying PTs in the \emph{grandcanonical}
ensemble by studying the distribution of complex zeros (DOZ) of the
grandcanonical partition function. Later on, Fisher \cite{Fi71} and
Grossmann et al. \cite{GrRo67,GrRo69,GrLe69}  employed a very similar
approach to analyze the temperature dependence of phase  
 transitions in the \emph{canonical} ensemble. Recently, further
significant progress in the understanding of critical phenomena has
been achieved by studying the connection between PTs and  phase
(or configuration) space topology \cite{CaCoPe02,TeSt04,Kastner04,FrPe04,FrPe05a,FrPe05b,AnEtAl05}.   
\par 
Formally, the seminal contributions
\cite{Ma37,MaAc37,MaHa38,StMa39,Eh12,YaLe52,LeYa52,GrRo67,GrRo69,GrLe69,Fi71} have
in common that, in the spirit of traditional thermodynamics, they
refer to macroscopically large systems; more exactly, to systems
satisfying the  thermodynamic limit (corresponding to $N,V,E\to
\infty$ such that number density  $n=N/V$ and energy density $e=E/N$ remain constant). However, the rapid experimental and computational progress during the
last two decades led to an increasing interest in extending 
thermodynamic concepts to \lq small\rq\space systems, containing -- by
definition -- only
a very limited number of
DOF~\cite{Hill,LePe61,WaBe94,WaDo95,Gr01}. Experiments on
finite systems include, e.g., investigations of two-dimensional Coulomb clusters in dusty plasmas \cite{KlMePiSc00,MeKlPi01}, Bose-Einstein condensation in magneto-optical traps
\cite{AnEnMaWiCo95,DaMeDrDuKuKe95}, and transitions in sodium 
clusters \cite{Sc01}. These experimental investigations were accompanied
by extensive theoretical and numerical studies (see, e.g.,
Refs.~\cite{BePe94,VaVlPeFo02,DaGiPiSt99}).    
\par  
Simultaneously, interest began to focus on the question
how to identify and classify the finite-size analogues of macroscopic
PTs. Import results in this regard were 
obtained by Wales et al. \cite{WaBe94,WaDo95}, who considered necessary and sufficient criteria for
phase coexistence in finite systems.  A general classification
scheme for smooth canonical PTs in small systems was proposed by 
Borrmann et al. \cite{BoMuHa00}. Pursuing an approach similar to that
of Yang and Lee \cite{YaLe52,LeYa52}, Fisher \cite{Fi71} and Grossmann et
al. \cite{GrRo67,GrRo69,GrLe69}, these authors suggest to use the DOZ
in order to characterize transitions in the \emph{canonical} ensemble
of small systems (cf. Sec. \ref{s:DOZ-scheme} 
below). M\"ulken et al. \cite{MuStBo01} and Alves et al. \cite{AlFeHa02} compare  the DOZ classification scheme with alternative proposals made by Gross \cite{Gr97,Gr01} and by Janke and Kenna \cite{JaKe01}, respectively. 
\par 
Such progress notwithstanding, there still  exist some open questions regarding which
types of non-analytic PTs can occur in small systems. For{\em
 (grand-)\-canonical} ensembles, it is well established  that truly
singular PTs can be observed in the thermodynamic  limit
only, corresponding to a system with formally infinite particle number
$N\to \infty$ \cite{YaLe52,LeYa52,Ru99,GrRo67,GrRo69,GrLe69,Fi71,BoMuHa00}. 
In contrast to this, the microcanonical thermodynamic functions (TDFs) may exhibit non-analytic
behavior even at finite $N$. For example, recently, 
Pleimling and Behringer \cite{PlBe05} have found singularities in 
microcanonical quantities of finite three-dimensional spin
models, which announce a continuous \emph{macroscopic} PT of the
infinite systems. Additionally, as we intend to demonstrate here by
means of very simple examples, the microcanonical 
TDFs of a small system can also exhibit non-analytic \emph{microscopic}
PTs\footnote{The appearance of such microcanonical singular
points even in the 1D case is \emph{not}
in conflict with van Hove's theorem for the canonical ensemble
\cite{Ho50,CuSa04}, since for most small systems (as well as for many \lq large\rq\space
systems) the microcanonical and canonical ensembles are generally \emph{not}
equivalent (see, e.g., Costeniuc et al. \cite{CoEtAl05}).}, characterized by well-defined critical energy values and typically
accompanied by strong variations of thermodynamic observables (e.g. oscillations of
temperature, heat capacity and pressure). From the physical point of
view, such microscopic PTs correspond to transitions between
different dissociation states of the system. Hence, they are important
indicators for essential structural changes in the small system under
consideration (quite analogous to singular points indicating
macroscopic PTs). 
For reasons of simplicity, the discussion in the present paper will be
restricted to one-dimensio\-nal (1D) models, but the general 
mechanism responsible for the singular and oscillatory behavior of 
microcanonical thermodynamic observables works analogously in two and three
spatial dimensions. Hence, it should in principle be possible to observe such 
microscopic oscillations in suitably designed evaporation/dissociation
experiments (which must, of course, realize the physical preconditions
of the microcanonical ensemble).    
\par
The paper is organized as follows: Section \ref{s:pt_micro} 
 is dedicated to microscopic PTs in the microcanonical ensemble (MCE) of
small model systems. As examples, we will 
consider isolated 1D chains with
Lennard-Jones (LJ) pair interactions and also the Takahashi
gas~\cite{DoGr72}. It will be shown that these simple systems exhibit  
singular microscopic PTs, separating different microcanonical
dissociation states. Subsequently, the Takahashi gas will be used in
Sec.~\ref{s:pt_canonical} to investigate the relation between singular
microscopic PTs in the MCE and smooth
PTs in the canonical ensemble (CE) as defined by the DOZ scheme \cite{BoMuHa00}. The paper concludes with a summary of the main results in
Sec.~\ref{s:summary}.

\section{Microscopic phase transitions in the microcanonical ensemble}
\label{s:pt_micro} 

Classical microcanonical thermodynamics refers to an ensemble of thermally isolated
systems, completely described by their Hamiltonian dynamics. Due to the fact that the systems are decoupled from the environment, the energy $E$ is a conserved quantity, i.e., there are no energy fluctuations in the MCE. In particular, as will be demonstrated in Secs. \ref{s:micro_example_LJ} and \ref{s:micro_example_3}, the microcanonical TDFs may exhibit singular points even in the case of small systems.  

\subsection{The microcanonical ensemble}
\label{s:micro} 

For the sake of simplicity we will confine ourselves to examples, where the
thermodynamic state is completely characterized by two control
parameters, namely, energy  $E$ and volume number $V$ (generalizations
to problems with additional macroscopic variables, e.g. the
strength of external magnetic fields, are straightforward). More
precisely, we will 
consider $N$ identical point-like particles of mass $m$, moving in $D$ equivalent spatial dimensions; i.e., the number of DOF reads $d=DN$ and the
volume is given by $V=L^D$, where $L$ is the length of the confining
cube (volume interval). The deterministic dynamics of the system is assumed to be
governed by a Hamiltonian of the standard form
\bse
\be\label{e:ham}
H(q,p;V)=K(p)+U(q;V)=E,
\ee
where $q=(q_1,\ldots, q_d)$ and  $p=(p_1,\ldots, p_d)$ are generalized
coordinates and momenta, respectively. The kinetic energy $K$ and the potential energy $U$ are given by
\be\label{e:EKin}
K(p)=\sum_{i=1}^d \f{p_i^2}{2m}\csp
U(q;V)=U_\mrm{pair}(q) + U_\mrm{box}(q;V), 
\ee
where $U_\mrm{pair}(q)$ represents the pair interactions of the particles, and the box potential is defined by 
\be\label{e:UBox}
U_\mrm{box}(q;V)=
\begin{cases}
0,& q\in [-L/2,L/2]^N,\\
+\infty,&  \mrm{otherwise}.
\end{cases}
\ee
\ese
The primary thermodynamic potential of the MCE is the entropy
$S=S(E,V)$, which is related to the microcanonical
\lq partition\rq\space function $\mcal{Z}_\mrm{M}(E,V)$ by 
\cite{Be67}   
\be\label{e:def-entropy}
S=k\ln\mcal{Z}_\mrm{M},
\ee   
with $k$ denoting the Boltzmann constant. The equations of state (EOS) for
the temperature $T$ and pressure $P$ are obtained by \cite{Hu63,Be67,Mu69,PeHaTi85,CaRa88,Gr01} 
\be\label{e:EOS}
\f{1}{T}\equiv\f{\p S}{\p E}\csp
\f{P}{T}\equiv\f{\p S}{\p V}.
\ee 
That is, the temperature $T$ of the MCE is a derived quantity, which is in
contrast to the CE (see
Sec. \ref{s:canonical} below), where the temperature is one of the
adjustable external control parameters.
\par
During the past century, various different
definitions for the microcanonical partition function $\mcal{Z}_\mrm{M}$, or the entropy $S$, have been proposed and investigated.
For classical systems as described by the Hamiltonian function
\eqref{e:ham}, the two most commonly used definitions for
$\mcal{Z}_\mrm{M}(E,V)$ read \cite{Hu63,Be67,Mu69,PeHaTi85,CaRa88,Gr01}   
\bse
\be
\mcal{Z}_\mrm{M}&=&\Go\label{e:def-omega},\\
\mcal{Z}_\mrm{M}&=&\eps_0\f{\p \Go}{\p E}
\label{e:def-Go-Gross},
\ee
\ese
where the phase volume $\Go$ is given by ($h$ denotes the Planck constant)
\be
\Omega(E,V)
\equiv\label{e:omega}
\f{1}{N!\,h^d}
\int_{\R^d}\diff q\int_{\R^d} \diff p \; \;
\Theta\bigl(E-H(q,p;V)\bigr).
\ee
The Heaviside unit step function $\Theta(x)$, appearing in Eq. \eqref{e:omega}, is defined by
$\Theta(x)=0$ for $x<0$ and $\Theta(x)=1$ for $x\geq0$.  
The additional parameter $\epsilon_0$ in  Eq. \eqref{e:def-Go-Gross} is a small
energy constant that quantifies the thickness of a thin energy shell around the
phase space surface defined by $H(q,p,V)=E$ and is formally required to make $\mcal{Z}_\mrm{M}$ dimensionless.
\par
It is well-known that the definitions  \eqref{e:def-omega} and
\eqref{e:def-Go-Gross}  may yield (almost) identical results
\cite{Ru99,Mu69,PeHaTi85,Be67} in the thermodynamic limit, i.e., if
$N$ is large. However, for small systems they lead to essentially different TDFs. To briefly illustrate this, let us consider an ideal gas with $N$ non-interacting  particles,
moving in the $D$-dimensional volume $V$. In this case, the phase volume is
given by \cite{Be67}
\be
\Omega(E,V)=
\f{(\pi\,2 mE)^{d/2}}{N!\;h^{d}\,\Gc(d/2 +1)}\,V^N,
\ee
where $\Gc(x)$ is Euler's Gamma-function.
Definition \eqref{e:def-omega} then yields the EOS
\be
E= \f{d}{2}\;k T\label{e:free-a}
\csp
\f{P}{T}=\f{k N}{V}\label{e:free-b},
\ee
whereas one obtains from definition \eqref{e:def-Go-Gross}
\be\label{e:Efree-false}
E=  \biggl(\f{d}{2}-1\biggr)\;k T
\csp
\f{P}{T}=\f{k N}{V}.
\ee
For systems with $d\gg 1$ the difference in the energy equations is negligible, but for small
systems it becomes relevant. In particular, for $d=1$
Eq. \eqref{e:Efree-false} yields a negative temperature at positive
energy. Obviously, a similarly unreasonable result is obtained for
$d=2$. This indicates that the partition function~\eqref{e:def-Go-Gross} is
 inappropriate for systems with low-dimensional phase space.  
\par
More generally speaking, only Eq. \eqref{e:def-omega} reproduces correctly
the well-known laws of thermodynamics and also yields the correct 
equipartition theorem for an arbitrary number $d$ of DOF,
whereas the Eq. \eqref{e:def-Go-Gross} leads to inconsistencies if $d$ is
small. This important aspect was first realized by Hertz \cite{He10,He10a}, and,  later on, also emphasized by Becker \cite{Be67}, Berdichevsky and v. Alberti \cite{Berdichevsky,BeAl91} and  Adib \cite{Ad02a}. In particular, denoting the ensemble average with respect to the
microcanonical probability density function $f(q,p)\propto
\gd(E-H(q,p))$ by $\langle\; \cdot\; \rangle_\mrm{MCE}$, it can be shown 
\cite{Be67,Hu63} that the (equipartition) identity  
\be
\f{k\,T}{2}=\f{1}{d}\langle K(p) \rangle_\mrm{MCE}
=\left\langle \f{p_i^2}{2m} \right\rangle_\mrm{MCE}\; \qquad \forall\; i=1,\ldots d
\ee
holds, only if one employs the Hertz entropy definition 
\be\label{e:def-entropy-2}
S&\equiv& k\ln \Go.
\ee 
Due to these reasons, all subsequent considerations will be based on Eq.~\eqref{e:def-entropy-2}.\footnote{This
is e.g. in contrast to Refs. \cite{Gr97,Gr01,MuStBo01} where the Boltzmann definition \eqref{e:def-Go-Gross} is considered.} For Hamiltonians as in
Eq. \eqref{e:ham} one can still perform the momentum integration in
Eq. \eqref{e:omega} using $d$-dimensional spherical coordinates, yielding
\be
\Omega(E,V)=\label{e:omega-1}
\f{O_d}{N!\,h^d\,d}\int_{\R^d}\diff q\;
\bigl\{2m\left[E-U(q;V)\right]\bigr\}^{d/2}\Theta\bigl(E-U(q;V)\bigr),
\ee
where $O_d={2\pi^{d/2}}/{\Gc({d}/{2})}$ denotes the surface of the $d$-dimensional unit sphere.
 
\subsection{Macroscopic vs. microscopic phase transitions}
\label{s:Ehrenfest_scheme}

Conventionally, \emph{macroscopic} PTs are singularities
(non-analyticities) of TDFs that arise in the thermodynamic limit,
corresponding to an infinitely large system. The first systematic classification of macroscopic PTs goes back to
Ehrenfest \cite{Eh12,Eh33}. According to the Ehrenfest scheme, a PT is
indicated by a non-analyticity of the Gibbs free enthalpy
$G\left(T,P,\ldots\right)$, assumed to be a function of the
temperature $T$, pressure $P$ and other external control parameters. 
The order of the PT is determined by the lowest order at
which any of the derivatives of $G\left(T,P,\ldots\right)$ becomes
non-continuous. Since the Ehrenfest scheme has turned out to be too
narrow in many cases, it is nowadays often preferred to merely distinguish \emph{discontinuous} (\emph{first-order}) and
\emph{continuous} (\emph{second-order}) transitions.
\par
Extending this concept to small systems, any non-analyticity of
the thermodynamic potential as function of the external control
parameters may be called a PT. However, to
avoid ambiguities, we shall speak of \emph{microscopic} PTs when
discussing singular (non-analytic) points in the microcanonical TDFs of small
systems. Furthermore, we will adopt
the following terminology to classify microscopic PTs in the
MCE: 
If the primary thermodynamic potential, the Hertz entropy $S$, is
discontinuous, then  we will call the PT discontinuous; if $S$ is 
non-analytic but continuous, the microscopic PT is called continuous.
\par
Let us next discuss how the formal (Ehrenfest-type) order of a microscopic PT
depends on the number of DOF. For systems described by Hamiltonian~\eqref{e:ham},  the phase volume \eqref{e:omega-1} is related to the admissible configuration space volume
\begin{equation}
\label{e:omega_config}
\omega(E,V)=\int_{\R^d}\diff q\;\Theta\bigl(E-U(q;V)\bigr)\;
\end{equation}
via\footnote{In an ordinary sense, Eq.~\eqref{e:omega-diff} is defined for even integer values $d>0$ only; however, by employing fractional derivatives \cite{LaOsTr76}, its range of validity can be extended to odd integer values $d>0$.}
\begin{equation}\label{e:omega-diff}
\f{\p^{d/2}\Omega(E,V)}{\p E^{d/2}}=\f{(2\pi\; m)^{d/2}}{ N!\,h^d}\;\omega(E,V)\;.
\end{equation}
If, at a given critical energy $E_c(V)$, $\omega(E,V)$ has continuous
derivatives up to order $j$, but a discontinuous $(j+1)$st
derivative, then $\Omega(E,V)$ and hence $S(E,V)$  have continuous
derivatives up to order $(j+d/2)$, but a discontinuity in the
$(j+d/2+1)$st derivative. Consequently, the formal order of the PT equals $(j+d/2+1)$, i.e., the order increases with an increasing number of DOF. For the
examples discussed below, the differentiability class of the
admissible configuration space volume $\omega(E,V)$ does not change
with particle number ($j=0$ for Lennard-Jones chains and $j=-1$ for
the Takahashi gas), and we indeed observe such an increasing  order with increasing particle number (cf. results of Sec.~\ref{s:micro_example_34}).\footnote{Recently, similar results have been
reported for the mean-field spherical spin model by Kastner and
Schnetz\cite{KaSc05}.} 

\subsection{Singular microscopic phase transitions in  
Lennard-Jones chains} 
\label{s:micro_example_LJ} 

To demonstrate the appearance of non-analytic microscopic PTs in the
MCE, we consider a 1D
Lennard-Jones (LJ) chain, moving freely in a 1D box volume $[-L/2,L/2]$. In this case, the pair potential in
Hamiltonian~\eqref{e:ham} reads:
\be\label{eq:LJN_Upair}
U_\mrm{pair}(q)= \f{1}{2}\sum_{\overset{i,j=1}{i\ne j}}^N U_\mrm{LJ}(|q_i-q_j|),
\qquad
U_\mrm{LJ}(r)=\label{a-e:LJ}4a
\left[\left(\f{\sigma}{r}\right)^{12}
-\left(\f{\sigma}{r}\right)^6\right],\qquad
\ee
where $a,\sigma>0$ are positive parameters. To simplify subsequent formulae, we will measure energy and length
in units of the parameters $a$ and
$r_0=2^{1/6}\sigma$, where $r_0$ is the position of the minimum of $U_\mrm{LJ}(r)$. With respect to these units the LJ potential 
\eqref{a-e:LJ} is given by 
\be
U_\mrm{LJ}(r)=\f{1}{r^{12}}-\f{2}{r^6}.
\ee
For small volumes $L\le1$ the LJ-force is
always repulsive, whereas in the more interesting case of
sufficiently large volumes, $L>1$, the LJ-force may also become
attracting.

\subsubsection{Diatomic Lennard-Jones molecule}
\label{s:micro_example_1} 

We start by discussing the simplest non-trivial example $N=2$ and $D=1$, where we
can calculate the microcanonical TDFs exactly.\footnote{The phase volume for the more complicated three-dimensional problem was recently calculated by
Umirzakov \cite{Um00}.} In this case, the energy $E$ can take values $E_0(L)\le E<\infty$, where the groundstate
energy is given by   
\be\label{a-e:E0}
E_0(L)\equiv\min_{(q,p)} H(q,p)=\begin{cases}
U_\mrm{LJ}(L),& L\le 1,\\
-1,&L>1.
\end{cases}
\ee 
\par
A straightforward calculation of the phase volume, based on relative and center-of-mass coordinates, yields  
\bse\label{a-e:pvol}
\be
\Go=
 \f{2\pi m}{h^2}\left(
\f{L}{11 r^{11}}-\f{1}{10 r^{10}}-\f{2L}{5r^5}+
\f{1}{2r^4}+Lr E-\f{r^2E}{2}\!\right)\biggl|_{r_\mrm{min}}^{r_\mrm{max}}\;,
\ee
where the boundary values are given by
\be\label{e:boundary}
r_\mrm{min}=X^{-1/6}\csp
r_\mrm{max}=\begin{cases}
L,& E\ge E_c(L)\equiv U_\mrm{LJ}(L),\\
Y^{-1/6},& E< E_c(L),
\end{cases}
\ee
\ese 
using the convenient abbreviations $X\equiv1+\sqrt{1+E}$ and $Y\equiv1-\sqrt{1+E}$.
\par
\begin{figure}[t]
\center
\epsfig{file=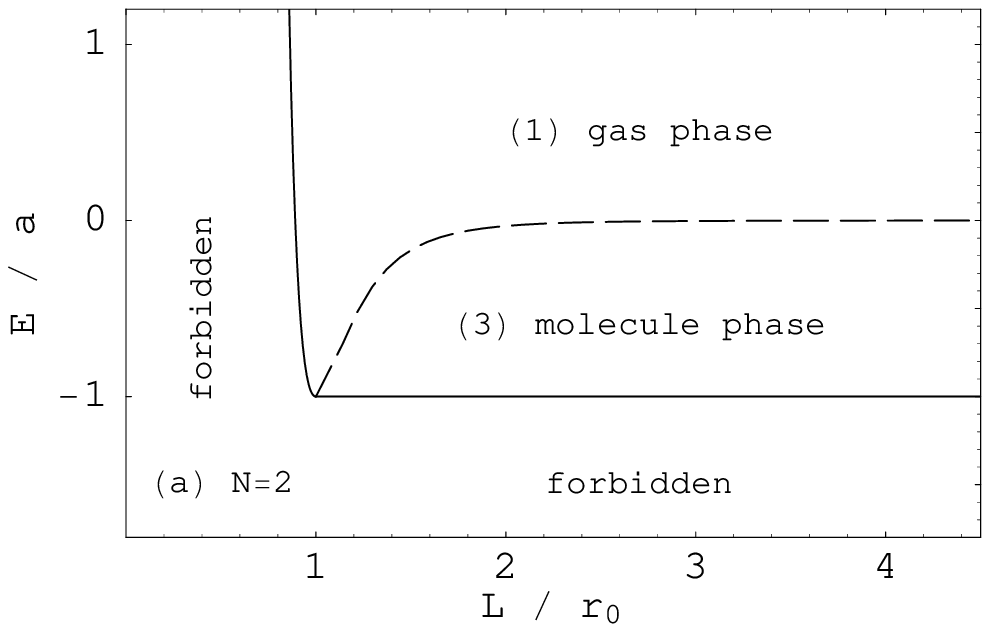 ,height=4.2cm, angle=0}
\epsfig{file=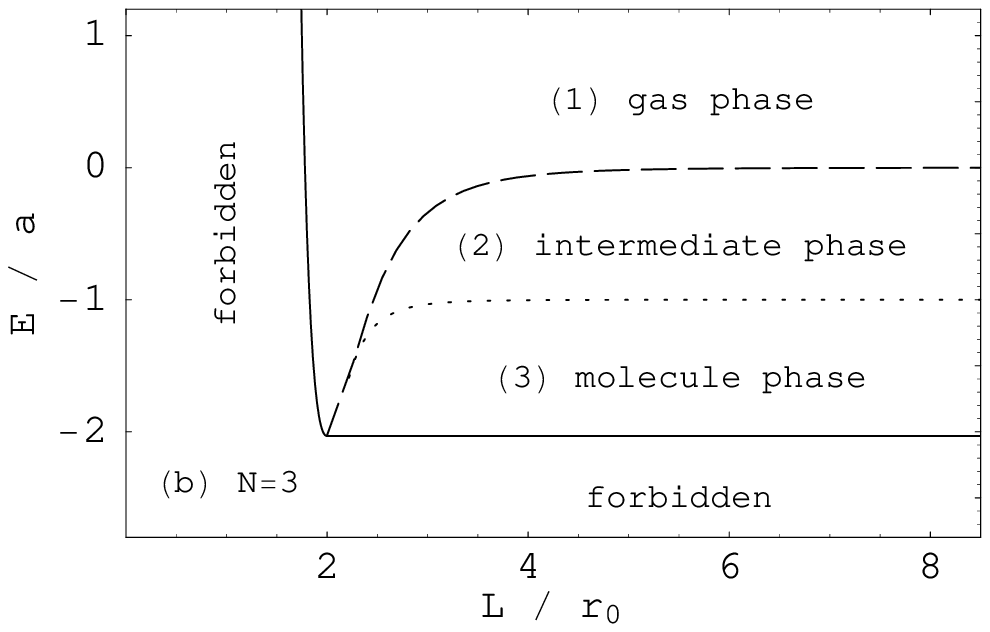 ,height=4.2cm, angle=0}
\caption{
\label{fig:LJ_phase_diagram}
Microcanonical phase diagrams for the 1D LJ
molecules. Energies below the minimum energy $E_0(L)$ (solid line) are
forbidden. (a) Diatomic LJ molecule ($N=2$): The critical curve
$E_{c}(L)$ (dashed line) separates a gas-like (or dissociated) phase
 from a molecule phase. (b) Triatomic LJ molecule ($N=3$): The
critical curve $E_{c1}(L)$ (dashed line) separates a gas-like (or
dissociated) phase from an intermediate
(partially bound) phase, enclosed by the critical curves  $E_{c1}(L)$ and $E_{c2}(L)$
(dotted line), and a bound molecule phase (3) below $E_{c2}(L)$. 
}
\end{figure}

In the case $L\leq1$ we have $r_\mrm{max}=L$ for all energies $E\geq E_0$, and, hence, the phase volume $\Go(E)$ is a smooth function for all permitted energy values.
For $L>1$, however, the boundary value $r_\mrm{max}$ changes its energy dependence at $E=E_c(L)$ in a non-analytic manner, and hence the phase volume $\Go(E)$ is not analytic at $E=E_c(L)$.
The critical curve $E_c(L)\equiv U_\mrm{LJ}(L)$, $L\geq1$ separates a gas-like phase
(dissociated state) from the  molecular phase (bound state) in the
$(L,E)$-parameter plane. This is illustrated in Fig.~\ref{fig:LJ_phase_diagram} 
and will become particularly evident from the expansions presented in the next
two paragraphs.
It is worthwhile to stress again that the critical curve $E_c(L)$ arises naturally due to the sudden change in the energy dependence of the phase volume, occurring when the energy $E$ passes the
critical curve.  
The microcanonical caloric curve $T(E)$ is continuous but not
differentiable along the critical transition curve $E_c(L)$, which is
located in the region of an S-bend or van der Waals-type loop,
respectively. Formally, this corresponds to a fourth-order transition.

\paragraph*{Super-critical energy values (dissociated phase).} 
Using the result \eqref{a-e:pvol} with $r_2=L$, corresponding to
supercritical energy values $E>E_c(L)$ -- or region \lq(1)\rq\space in
Fig. \ref{fig:LJ_phase_diagram}, respectively -- we can derive from
Eqs. \eqref{e:EOS} the microcanonical EOS, yielding
\bse\label{a-e:EOS-gas}
\be
k T&=&
\f{Z+
\f{24 L (X-5E)}{X^{1/6}}-\f{33(X-2E)}{X^{1/3}}}
{55\left
[L^{2}-\f{(E+X)(2LX^{1/6}-1)}{X^{4/3}(1+E)^{1/2}}\right]},
\\
\f{p L}{k T}&=&
\f{110L^2E+\f{10}{L^{10}}-\f{44}{L^{4}}-
\f{2 L \left[X(5X-22)+55E\right]}{X^{1/6}}}
{Z +
\f{11\left[X(X-5)+5E\right]}{X^{1/3}}
-\f{2L \left[X(5X-22)+55E\right]}{X^{1/6}}},
\ee
\ese
where $ Z\equiv 55L^2E+11L^{-4}-L^{-10}$. Taking the high-energy limit at constant volume $V=L$ one finds
\be
\lim_{E\to \infty} \f{k T}{E} =1
\csp
\lim_{E\to \infty} \f{PL}{k T} = 2\;,
\ee 
corresponding to the laws for the ideal 1D
two-particle gas. Hence, the parameter region $E>E_c(L)$ can be
identified as two-particle gas state or dissociated phase, respectively.  

\paragraph*{Sub-critical energy values (bound phase).} 
Because of Eq. \eqref{a-e:E0}, the opposite case $E< E_c(L)$ --
corresponding to region \lq(3)\rq\space in
Fig. \ref{fig:LJ_phase_diagram} (a) --
can only be realized, if $L>1$ holds; i.e., if the box volume is larger than
the distance corresponding to the potential minimum $r_0$. Again using 
Eqs. \eqref{a-e:pvol}, but this time with $r_2=Y^{-1/6}$, we obtain 
\bse\label{a-e:EOS-mol-T}
\be
k T&=&
\f{3\left[8L\,g(X,Y)-11\,f(X,Y)\right](-E)^{1/3}}
{55\left[X^{1/3}-Y^{1/3}+2LX^{1/6}(Y^{1/3}-(-E)^{1/6})\right]},
\\
\f{p L}{k T}&=&
\f{8L\, g(X,Y)}
{8L\,g(X,Y)-11\,f(X,Y) }
\label{a-e:EOS-mol-p},
\ee
\ese
where we have made use of the abbreviations
$$
g\equiv(1+5X)Y^{5/6}-(1+5Y)X^{5/6},
\qquad
f\equiv(1+2X)Y^{2/3}-(1+2Y)X^{2/3}.
$$
Expanding EOS \eqref{a-e:EOS-mol-p} near the groundstate energy $E_0=-1$ yields
\bse
\be
\label{a-e:exp-T}
k T&=&\f{2}{3}(E+1)+\mcal{O}\left[(E+1)^{2}\right],\\
\label{a-e:exp-p}
\f{P}{k T}&=&\f{1}{L-1}+\mcal{O}\left[(E+1)^{1}\right].
\ee
\ese  
Equation \eqref{a-e:exp-T} indicates that, at low energy, each momentum variable as
well as an approximately harmonic excitation of the relative coordinate
carries on average the energy amount  $kT/2$. Obviously, this is in
agreement with the 
equipartition theorem for harmonic DOF. Equation \eqref{a-e:exp-p} corresponds to the pressure law
for an ideal one-particle gas in the reduced 1D volume
$V_\mrm{eff}=L-1$, reflecting the fact that, at sufficiently low
energy values, the two particles form a bound molecule with distance $r\approx
1$ between each other.  
\par
\begin{figure}[t]
 \epsfig{file=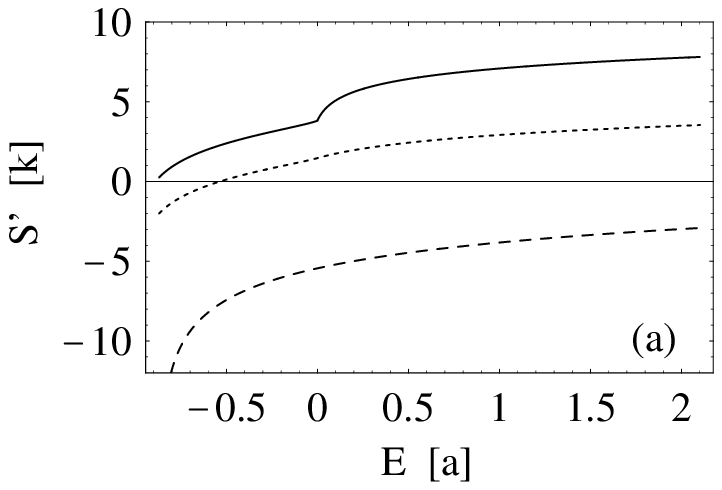 ,height=4.2cm, angle=0}
 \epsfig{file=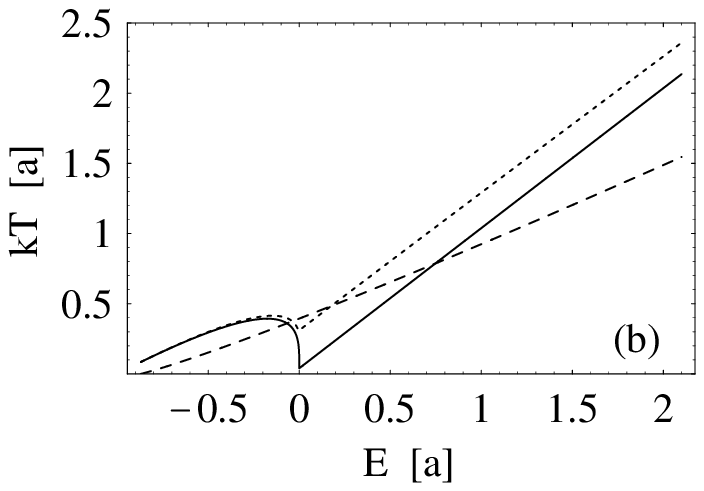 ,height=4.2cm, angle=0}
 \caption{Microcanonical TDFs for the
 1D diatomic LJ molecule ($N=2$). Energy is measured in units of the binding energy $a$. Volume $V=L$ is measured in units of the parameter $r_0$,
 corresponding to the minimum of the LJ potential. The mass unit is
 chosen such that $m=1$. (a) Hertz entropy $S'=S-S_0$
 as function of the energy for $L=20$ (solid line),  $L=3$ (dotted 
 line), and  $L=0.95$ (dashed),  where  
 $S_0=-k\ln[h^2/(mr_0^2 a^2)]$. Note the convex curvature of the solid
 curve at the transition energy. (b) Caloric curves for $L=20$ (solid line),  $L=3$ (dotted 
 line), and  $L=0.95$ (dashed). One can readily see the
 singularity (peak) in the S-bend region, occurring exactly when the critical
 line  $E_c(L)=U_\mrm{LJ}(L)$ in Fig. \ref{fig:LJ_phase_diagram} is
 crossed. 
 \label{fig02}}
 \end{figure}

In Fig. \ref{fig02} (b)  the caloric curves and the
pressure law are shown for different fixed values of $L$ and $E$,
respectively. When the volume is large enough, $L\ge 1$, the caloric
curves exhibit a characteristic convex region (S-bend) and, in particular,
also a non-differentiable point (see solid and dotted
curves). These kinks occur when the energy
passes through the critical value $E_c(L)$.


\subsubsection{Lennard-Jones chains with $N>2$ particles}
\label{s:micro_example_34} 

Analogous microscopic PTs do also occur in LJ molecules with larger particle numbers. For $N>2$ it is very difficult or, perhaps, even impossible to express
the phase volume~\eqref{e:omega-1} in terms of closed
functions. Usually, one can perform only one or two of the $N$
integrations analytically, and the remaining integrals have to be
calculated numerically, using e.g. Monte-Carlo methods. We employed
the Divonne algorithm of the {\scshape Cuba} library~\cite{CubaLib} to calculate the phase volume for LJ molecules with 3 and 4 particles. We used at least 1 million sample points
and partially increased the sample size up to 100 million points for
testing. We also cross-checked the results with other
deterministic and probabilistic integration algorithms of the
{\scshape Cuba} library and found excellent agreement.

\begin{figure}[b]
\epsfig{file=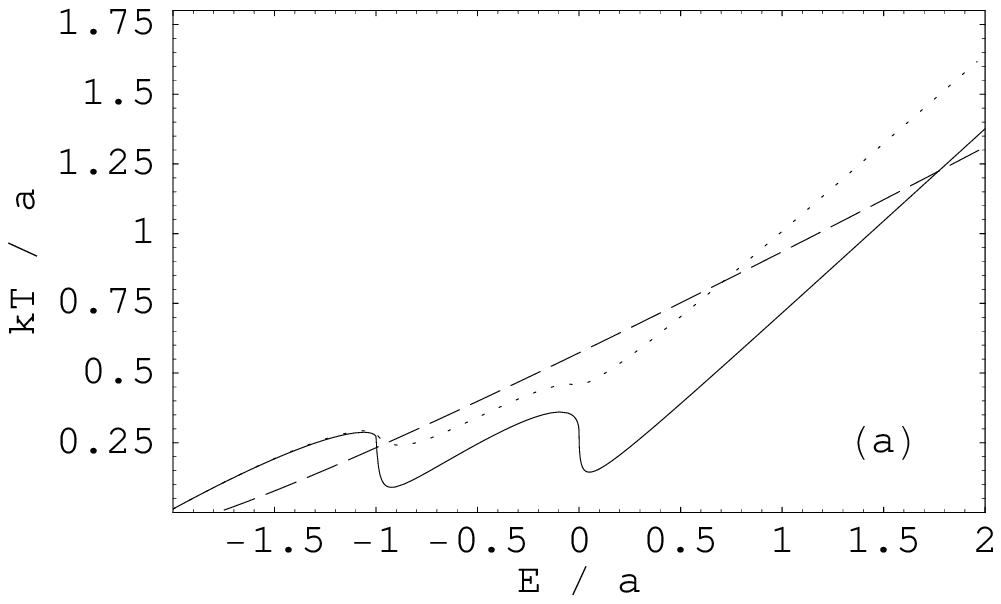 ,height=4.2cm, angle=0}
\epsfig{file=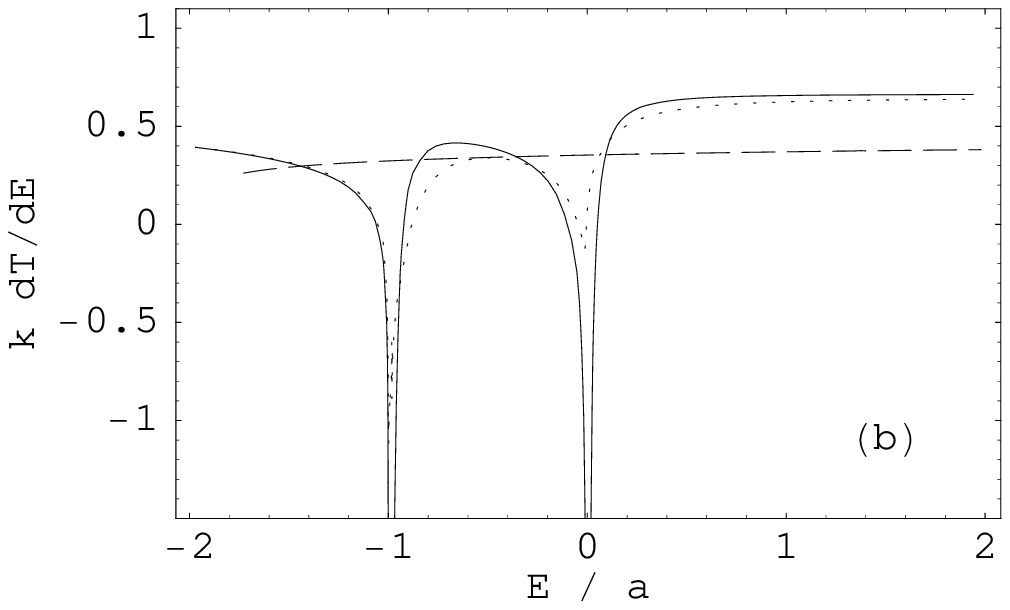 ,height=4.2cm, angle=0}
\caption{
\label{fig03}
LJ chain with $N=3$ particles. (a) The microcanonical caloric curve
$T(E)$, and (b) its first derivative for different values of the volume $L=1.9$ (dashed
line), $L=6$ (dotted line), and $L=40$ (solid line). $T(E)$ shows two
S-bend regions for $L=6$ and $L=40$. In these regions, the first
derivative $\diff T(E)/\diff E$ exhibits a lambda peak pointing
downward (the second derivative $\diff^2T(E)/\diff E^2$ has a pole). 
}
\end{figure}

Figure~\ref{fig03} shows the numerically calculated microcanonical caloric
curve $T(E)$ of the three-particle LJ chain ($N=3$) and its first derivative for different
values of $L$. For $L=1.9$, the caloric curve appears smooth and
almost linear. For $L=6$ and $=40$, $T(E)$ is still continuous, but
shows two S-bend regions around energies $E_{c1}\approx-1$ and
$E_{c2}\approx 0$. At these energies, the first derivative of $T(E)$
exhibits a (negative) `lambda peak', whereas the second derivative
diverges. 
\par
As in the two-particle case, the phase volume $\Omega(E,L)$ can be written as an integral with an integrand which is analytic for all values of $E$ and $L$, and integration boundaries which are analytic except for certain values of $E$ and $L$. A detailed analysis yields the two critical curves
\begin{align}
\notag
E_{c1}(L) &= 
U_\mrm{LJ}\left(\f{L}{2}\right)+U_\mrm{LJ}\left(\f{L}{2}\right)
+U_\mrm{LJ}\left(L\right)
\;&\text{if}\; L\geq \left(\f{2731}{43}\right)^{1/6},\\
 E_{c2}(L) &= \notag
U_\mrm{LJ}\left(r_{c2}(L)\right)+U_\mrm{LJ}\left(L-r_{c2}(L)\right)+U_\mrm{LJ}\left(L\right)&\text{if}\; L\geq2\left(\f{13}{7}\right)^{1/6}\;,
\end{align}
where $r_{c2}(L)$ is given by a polynomial equation of degree eighteen, with $r_{c2}(L)\approx-1$ for \mbox{$L\gg1$}.
The resulting phase diagram is shown in
Fig.~\ref{fig:LJ_phase_diagram} (b). One readily identifies three microscopic 
phases separated by the critical curves $E_{c1}(L)$ and
$E_{c2}(L)$. 
\par
At high energies $E>E_{c1}(L)$, the system is in a gas-like, fully dissociated phase. The relative positions $r_{21}=|q_2-q_1|$ and $r_{32}=|q_3-q_2|$ of the
particles are only restricted by the hard-core repulsive part of the
interaction potential, but apart from this constraint the particles can move independently inside the remaining volume. In the high-energy limit, one finds
$E\approx \f{3}{2} kT$, corresponding to a quasi-ideal 1D three-particle gas.
\par
For $L>2\left(\f{13}{7}\right)^{1/6}$ and low energies $E_0(L)<E<E_{c2}(L)$, the system is in a bound molecule phase. The relative positions  $r_{21}$ and $r_{32}$ are restricted by the interaction potential to be close to the equilibrium position. 
\par
For $L>2\left(\f{13}{7}\right)^{1/6}$ and intermediate energy values $E_{c2}(L)<E<E_{c1}(L)$, the system is in a partially dissociated phase. One of the relative positions  $r_{21}$ and $r_{32}$ is
restricted to be close to its equilibrium value, whereas the other is
only restricted by the hard-core repulsive part of the interaction
potential and the box volume. Accordingly, one of the three particles may move rather independently inside the box, whereas the other two remain bound to each other.

Although more complicated in detail, the calculations for the four-particle LJ chain are in principle the same as for $N=3$. We briefly list the main results:
For $N=4$ (and only taking nearest-neighbour interaction into account), three critical lines $E_{c1}(L)$, $E_{c2}(L)$ and $E_{c3}(L)$ divide the $(L,E)$-plane into four different phases. At
high energies or at low volumes $L<L_0$, $L_0=3$, the system
is in a gas-like phase. At low energies and large volume, the system
is in a molecule phase, where the particles can only move as a whole
molecule inside the volume. Between the gas phase and the molecule
phase, there are now \emph{two} intermediate, or partly dissociated,
phases, where the LJ molecule is broken up into two or three parts
consisting of one or two particles. For large volumes, the 
caloric curve $T(E)$ shows three PTs with
a continuous first, but discontinuous second derivative. This confirms that the formal order of the microscopic PTs increases with
particle number as discussed in Sec.~\ref{s:Ehrenfest_scheme}.

\subsection{Takahashi gas}
\label{s:micro_example_3} 

As another example, let us consider the piecewise linear pair potential 
\be\label{e:upair}
u(r)=
\begin{cases}
+\infty,& |r|=0,\\
a |r|/r_0,& 0<|r|\le r_0\\
1,&|r|>r_0
\end{cases}
\ee
where  $a$ and $r_0$ are positive parameters. In the literature this
model is known as the Takahashi gas \cite{DoGr72}. The potential $u$ from Eq. \eqref{e:upair} is qualitatively similar to the LJ potential, but, unlike $U_\mrm{LJ}$, it allows to calculate the exact TDFs for both MCE and CE in the case $N=2$. The Takahashi gas is, therefore, particularly well suited for studying the differences between the two ensembles.
\par
To keep subsequent calculations as simple as possible, we will
measure energy and length in units of $a$ and $r_0$ from now on ($r_0$
now defines the range of the potential). With respect
to these units, the potential in  Eq. \eqref{e:upair} simplifies to 
\be\label{e:upair-rescaled}
u(r)=
\begin{cases}
+\infty,& |r|=0,\\
|r|,& 0<|r|\le 1,\\
1,&|r|>1,
\end{cases}
\ee
and, in contrast to Eq. \eqref{a-e:E0}, the groundstate energy is now
given by \mbox{$E_0(L)=0$}. A straightforward calculation of the phase volume yields:
\be
\Go
= \label{e:pvol-3}
\f{\pi m}{3h^2}\;
\bc
L^2(3E-L)     ,&E> E_c(L),\quad L\le 1,\\
3L[L(E-1)+1]-1        ,&E> E_c(L),\quad L >1, \\
E^2(3L-E) ,&E\le E_c(L),
\ec
\ee
where the critical energy curve is given by $E_c(L)\equiv u(L)$.
Note that, compared with LJ potential from
Sec. \ref{s:micro_example_1}, we must now additionally distinguish
the cases $L\le 1$ and $L>1$, but this is only because the piecewise linear
potential is not differentiable at $r=1$. However, analogous to the case of the LJ potential, the critical curve $E_c(L)=u(L)$ separates a  gas-like phase (dissociated state) from the molecule phase (bound state) in the
$(L,E)$-parameter plane. This becomes evident from the microcanonical
EOS:
\bse\label{e:EOS-gas}
\be
\f{1}{k T}&=&
\bc
\f{3}{3E-{L}}     ,&E> E_c(L),\quad L\le 1,\\
\f{3L^2}{3L[L(E-1)+1]-1}    , &E> E_c(L),\quad L >1, \\
\f{2}{E}+\f{1}{E-3L} ,&E\le E_c(L).
\ec\\
\f{P}{k T}&=&
\bc
\f{3L-6E}{L^2-3LE}     ,&E> E_c(L),\quad L\le 1,\\
\f{6L(E-1)+3}{3L[L(E-1)+1]-1},           &E> E_c(L),\quad L >1, \\
\f{3}{3L-E} ,&E\le E_c(L).
\ec
\ee
\ese
Taking the high-energy limits of
Eqs. \eqref{e:EOS-gas} at constant volume $V=L$ one finds 
\be
\lim_{E\to \infty} \f{E}{k T} =1
\csp
\lim_{E\to \infty} \f{PL}{k T} = 2\;,
\ee 
corresponding to the laws for the ideal 1D
two-particle gas. Hence, the parameter region $E>E_c(L)$ can be referred to as
 gas-like or dissociated phase, respectively.
\par
In the opposite case, $E\le E_c(L)$, a low-energy expansion of the EOS near the groundstate energy $E_0=-1$ yields
\be
\label{e:exp-p}
\f{P}{k T}&=&\f{1}{L}+\mcal{O}\left(E^{1}\right).
\ee
Neglecting terms of higher order, Eq. \eqref{e:exp-p} corresponds to the
pressure law for an ideal one-particle gas. This reflects that at sufficiently low energy the two particles form a bound
LJ-type molecule. Compared with the results of the preceding
section, a slight difference is given by
the fact that in the LJ case the particles have a non-vanishing
distance in the groundstate, see Eq.~\eqref{a-e:exp-p}.
\par
\begin{figure}[t]
 \center 
 \epsfig{file=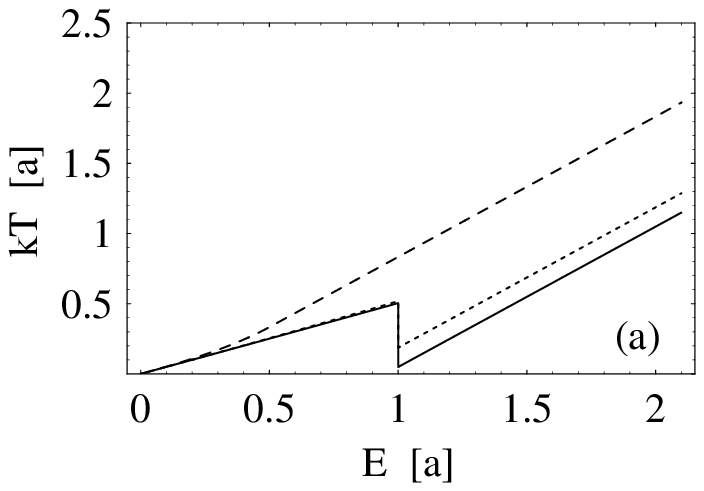 ,height=4.2cm, angle=0}
 \epsfig{file=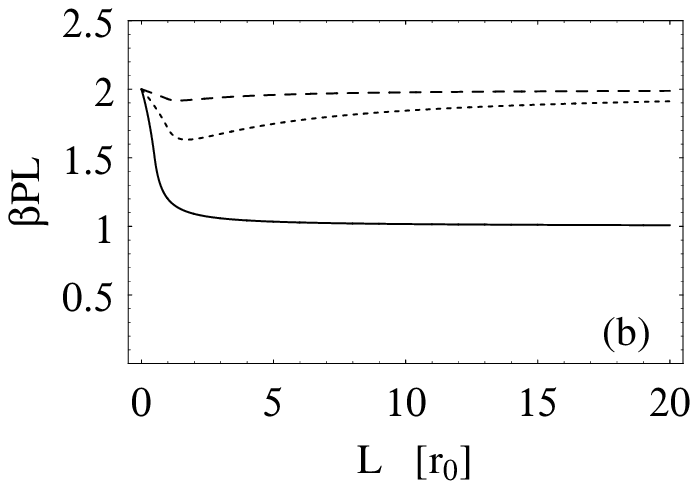 ,height=4.2cm, angle=0}
 \caption{Microcanonical TDFs for the
 1D diatomic ($N=2$) Takahashi gas. (a) Caloric curves for $L=20$ (solid line),  $L=5$ (dotted), and  $L=0.5$ (dashed). One
 can see the jump singularity for $L>1$, occurring exactly  when the critical
 line  $E_c(L)=u(L)$ is crossed. (b) Isoenergetic pressure curves for $E=0.5$
 (solid line),  $E=1.5$ (dotted), and  $E=5.0$ (dashed), where $\gb\equiv(kT)^{-1}$. The local minimum
 is observable only if $E/a\gtrsim 1$. In the limit $L\to \iy$
 the particles behave as an ideal two-particle (one-particle) gas, if
 they are in the dissociated phase $E>a$ (bound phase $E<a$).
\label{fig04}}
 \end{figure}

In Fig. \ref{fig04} (a) and (b) the caloric curves and the
pressure law are shown for different fixed values of $L$ and $E$,
respectively. In particular, in Fig. \ref{fig04} (a) one can see that the
caloric curves exhibit a singularity, when the energy passes through the
critical value $E_c(L)$.  More exactly, for $L\le 1$ the caloric curves are
continuous but not differentiable at $E=E_c(L)$, corresponding to a
third-order PT. By contrast, in the complementary case
$L>1$, the caloric curves become discontinuous at $E_c(L)$,
corresponding to a second-order PT (which is a
consequence of the additional singularity at $r=1$ and the vanishing gradient of the piecewise linear potential at $r>1$). 
\par
In the next section, we are going to study the relationship between
this singular microcanonical PT and the critical behavior of the corresponding systems in the CE. 

\section{Smooth phase transitions in the canonical ensemble}
\label{s:pt_canonical} 
Although referring to different physical conditions, MCE 
and CE may yield (almost) identical TDFs for
well-behaved 
systems with a large number of DOF $d\to
\infty$. However, this \lq equivalence\rq\space between the different
statistical ensembles does usually not hold for systems with a small number of
DOF \cite{Be67,Gr01,NoltingBand6,LaLi5,CoEtAl05}. Hence, one has to describe
the thermodynamics of small systems by the statistical ensemble which
actually corresponds to the given physical conditions.
 
\subsection{The canonical ensemble}
\label{s:canonical} 

Considering the CE is appropriate if
the system under investigation is coupled to an infinite heat bath
\cite{Be67,Ch43,HaTaBo90,Kampen}. The surrounding heat bath (thermostat) keeps the
temperature of the particles constant, but causes energy fluctuations
$\delta E>0$ around the energy mean value $\bar{E}$. Consequently, 
singularities in the thermodynamic functions may only exist in the thermodynamic limit with $N,E\to
\infty$, such that $e=E/N$ and $n=N/V$ remain constant and $\delta e
\to 0$; i.e., for a finite system
with $N<\infty$ the heat bath smoothens the singularities due to
non-vanishing fluctuations $\delta e$. Nevertheless, it is possible to
define and to classify \lq smooth\rq\space PTs by
virtue of the DOZ scheme \cite{BoMuHa00}, discussed below.
\par
Given a Hamiltonian of the form \eqref{e:ham}, the canonical partition
function is defined by 
\be\label{e:canonical_partition_function}
\mcal Z_\mrm{C}(\gb,V)=
\f{1}{N!h^d}\int_{\R^d}\diff q\int_{\R^d} \diff p \; \;
\exp[-\gb H(q,p;V)],
\ee
where $\gb\equiv(kT)^{-1}$. The external control variables are now $(T,V)$ or $(\gb,V)$, respectively. When 
$\mcal Z_\mrm{C}(\gb,V)$ is known, (mean) energy and pressure of the
CE are obtained by differentiation    
\be\label{e:CE-EOS}
\bar{E}\equiv
-\f{\p}{\p\gb}\ln \mcal{Z}_\mrm{C}\csp
\bar{P}\equiv
-\f{\p F}{\p V}
\ee
where $F\equiv -k T \ln \mcal{Z}_\mrm{C}$ is the free energy. For convenience, we are going to drop the over-bars
and simply write $E$ and $P$ in the next section (over-bars will be reinstated occasionally, e.g., when comparing
microcanonical and canonical quantities).  

\subsection{Classification of canonical phase transitions in the DOZ scheme} 
\label{s:DOZ-scheme}

According to Yang and Lee \cite{YaLe52,LeYa52}, the
distribution of zeros (DOZ) of the grandcanonical partition function
determines the PTs in the grandcanonical ensemble. Later
on, a similar approach
has been employed by Fisher \cite{Fi71} and Grossmann and Rosenbauer \cite{GrRo67} to
identify and classify PTs in the canonical
ensemble. They considered the distribution of complex zeros
$\tilde{\beta}_k$  of the canonical partition function
$\mcal{Z}_\mrm{C}(\tilde{\beta})$, taken as complex function of the
complex inverse temperature 
$$
\tilde{\beta}=\beta+i\tau\hspace{1em}(\beta>0)\;.
$$ 
For finite systems, one can show that there are no zeros on the
positive real
axis, i.e., $\Im(\tilde{\beta}_k)\neq0$ $\forall k$. In the
thermodynamic limit, however, certain points $\beta_c$ on the real 
$\beta$-axis may become limiting values of the DOZ. By studying how the zeros condense near the $\beta$-axis, one can characterize the PT. 
\par 
Although for small systems there are, strictly speaking, no non-analytic PTs in the CE, it may nevertheless be helpful to distinguish different thermodynamic phases~\cite{BePe94,VaVlPeFo02}. The DOZ classification scheme of Bohrmann et al.~\cite{BoMuHa00} is based on the idea that the complex zeros 
$\tilde{\beta}_k$ closest to the real $\beta$-axis can be employed
to estimate the DOZ behavior in the thermodynamic limit. The 
extrapolated limiting values  $\beta_c$ are used to define the
`smooth' canonical PT of the finite system.  To be more specific, one first numbers  the complex zeros $\tilde{\beta}_k$, $k=1,2,\ldots$ of the canonical
partition function $\mcal{Z}_\mrm{C}(\tilde{\beta})$ according to
their distance to the real $\beta$-axis, and then calculates the quantities
\bse\label{e:Borrmann}
\be
\gamma&=&\f{\beta_2-\beta_1}{\tau_2-\tau_1},\\
\phi_k&=&\f{1}{2}\left(\f{1}{\left|\tilde{\beta}_{k} -
\tilde{\beta}_{k-1}\right|} +
\f{1}{\left|\tilde{\beta}_{k+1} - \tilde{\beta}_{k}\right|}\right),\qquad
k=2,3,\ldots\;,\\
\alpha&=&\f{\log(\phi_3)-\log(\phi_2)}{\log(\tau_3)-\log(\tau_2)},\\
\beta_c&=&\beta_1-\gamma\tau_1.
\ee
\ese
These quantities characterize the DOZ near the real $\gb$-axis. According to the scheme of Borrmann et al. \cite{BoMuHa00}, a first-order
transition at temperature $T_c=(k\gb_c)^{-1}$ appears if $\alpha=0$ and
$\gamma=0$, whereas values  $1>\alpha>0$  correspond to second-order
transitions, and $\ga>1$ to even higher-order transitions in the
original Ehrenfest classification. In order to actually observe a singular PT in the Ehrenfest sense, it is required that $\tau_1\to 0$
in the thermodynamic limit. 
\par
A similar classification scheme for smooth canonical PTs in small systems,
based on the average cumulative density of zeros of the canonical
partition function, has been proposed by Janke, Kenna et al.  
\cite{JaKe01,JaJoKe04}. Alves et al. \cite{AlFeHa02} compared both approaches
for a variety of model systems.   
In the following, however, we will refer to the DOZ-scheme by Borrmann et
al. \cite{BoMuHa00} in order to classify PTs in the
CE of a small system. In particular, we aim to compare
the DOZ-classification with the MCE results obtained for the Takahashi gas. 

\subsection{Takahashi gas}
\label{s:canon_example_2} 

We first calculate the canonical TDFs for the Takahashi model from Sec. \ref{s:micro_example_3},
corresponding to two point-like particles, $N=2$, confined in a
1D volume $V=L$ and interacting via the (rescaled) piecewise linear pair potential from Eq. \eqref{e:upair-rescaled}.  For $L\le 1$, we find explicitly
\bse\label{e:Z_canon}
\be
\mcal{Z}_\mrm{C}
&=&\label{e:Z_canon-a}
2\;\f{m\pi}{h^2\gb^3}\; 
\left(e^{-\gb L}+\gb L-1\right),
\ee  
and in the opposite case, that is for $L>1$,
 \be
\mcal{Z}_\mrm{C}
&=&\label{e:Z_canon-b}
2\;\f{m\pi}{h^2\gb} \left\{
\f{1}{\gb^2}\left[e^{-\gb}\left(1+\gb-\gb L\right)+\gb L-1\right]
+\f{e^{-\gb}}{2}(L-1)^2
\right\}.  
\ee
\ese
Accordingly, one obtains the following canonical EOS
\bse\label{e:EOS_canon}
\be\label{e:EOS_canon-a}
E&=&
\bc
\f{3+\gb L+e^{\gb L} (2\gb L-3)}
{\gb[1+ e^{\gb L}(\gb L-1)]},
&L\le 1,\\
\f{6\gb +(6-4\gb L)(1-e^\gb)+\gb^2(L-1)[L-3+\gb(L-1)]}
{\gb\{1+[1-\gb(L-1)]^2+2e^\gb(\gb L-1)\}},
&L> 1.
\ec\\
P&=&
\bc
\f{e^{\gb L-1}}{e^{\gb L}(\gb L-1)+1},
&L\le 1,\\
\f{2[e^\gb-1+\gb(L-1)]}
{1+[1-\gb(L-1)]^2+2e^\gb(\gb L-1)}
&L> 1,
\ec
\ee
\ese
In the high-temperature limit, corresponding to $\gb\to 0$ at constant volume $V=L$, one finds
\be
\lim_{\gb\to 0} \gb E =1
\csp
\lim_{\gb\to 0} \gb PL = 2\;,
\ee 
i.e., the system behaves like an ideal 1D
two-particle gas in this limit. 
\begin{figure}[h]
 \epsfig{file=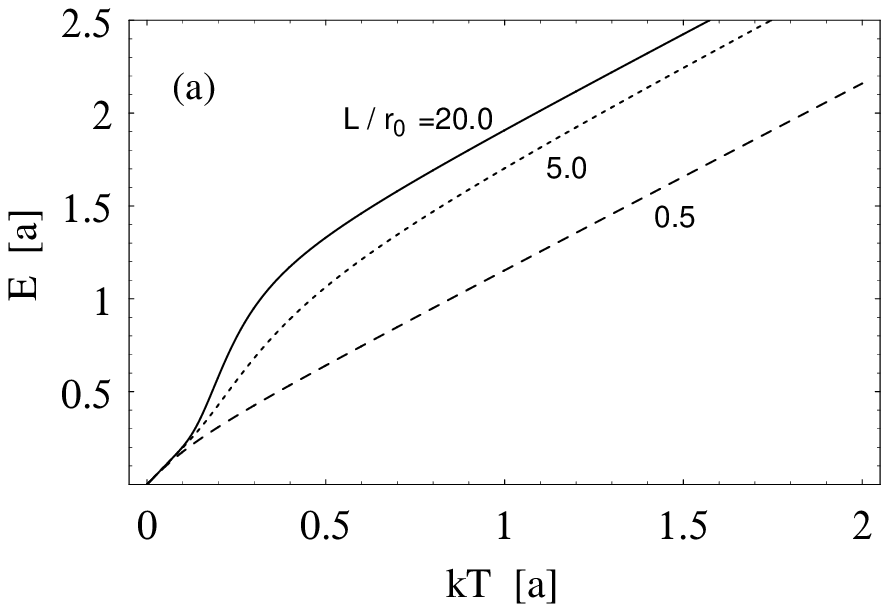 ,height=4.2 cm, angle=0}
 \epsfig{file=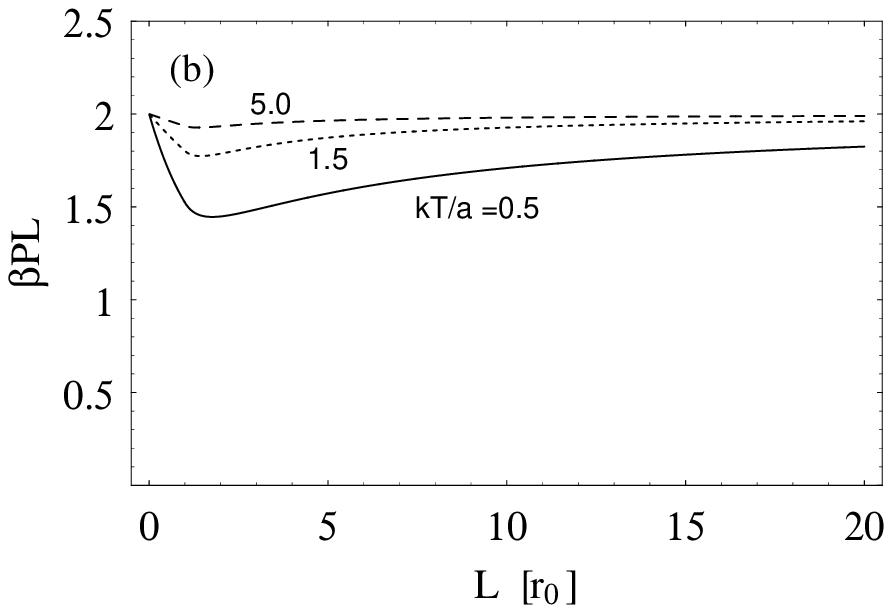 ,height=4.2 cm, angle=0} 
 \caption{Canonical TDFs for the
 1D diatomic Takahashi gas. Mean energy $E$ and
 thermal energy $kT$ are given in units of the binding energy $a$. The volume $V=L$ is measured in units of the parameter $r_0$,
 corresponding to the range of the potential, and the mass unit
 is chosen such that $m=1$ holds. (a) Caloric curves for  different
 fixed  values of $L$.  (b) Pressure law for different fixed
 temperature values $kT$.  Note, that $\gb PL\to 2$ for $L\to \iy$, 
 corresponding to the law of the ideal two-particle gas.  
\label{fig05}}
 \end{figure}
\par
\begin{figure}[h]
 \centering 
 \epsfig{file=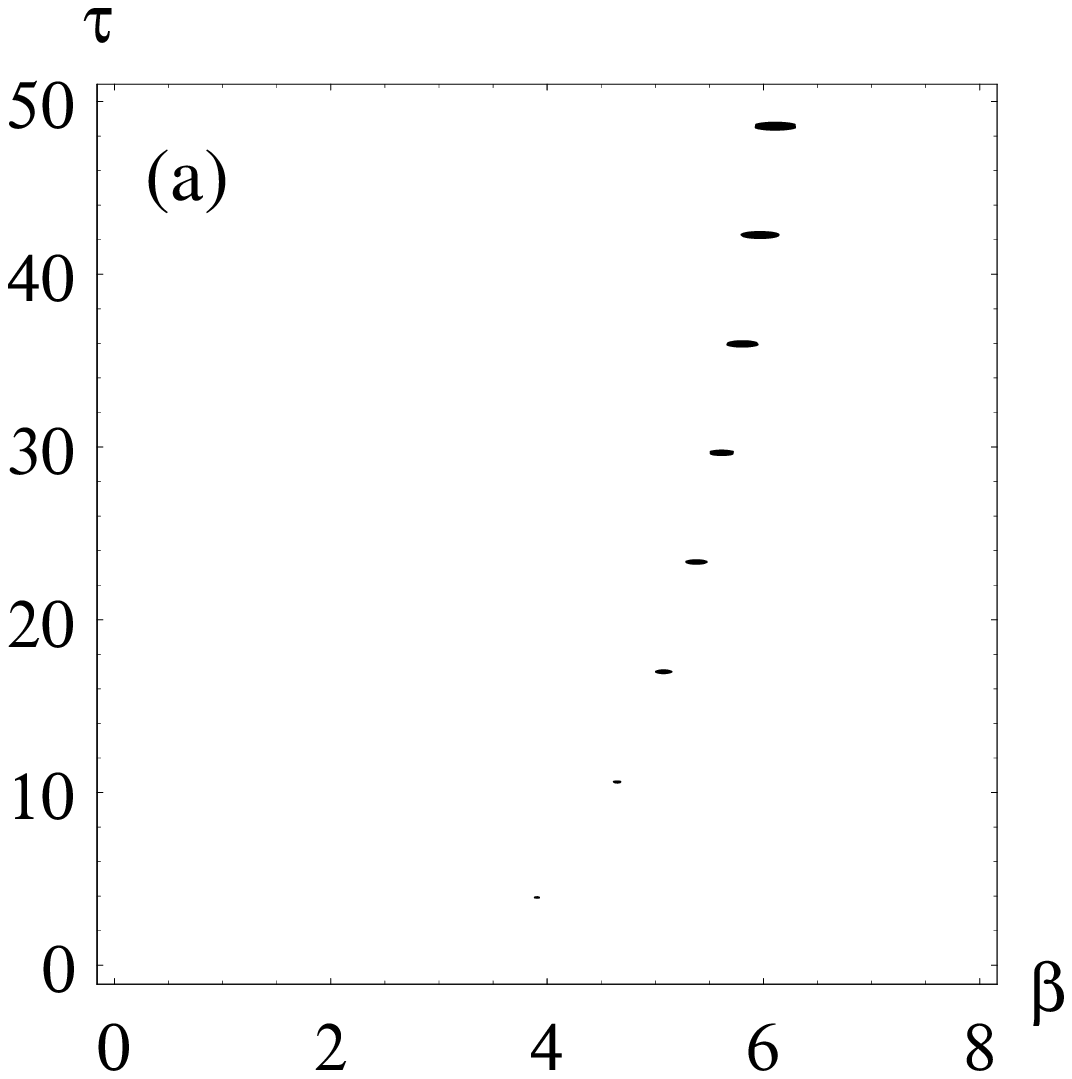 ,height=5.5 cm, angle=0}
 \epsfig{file=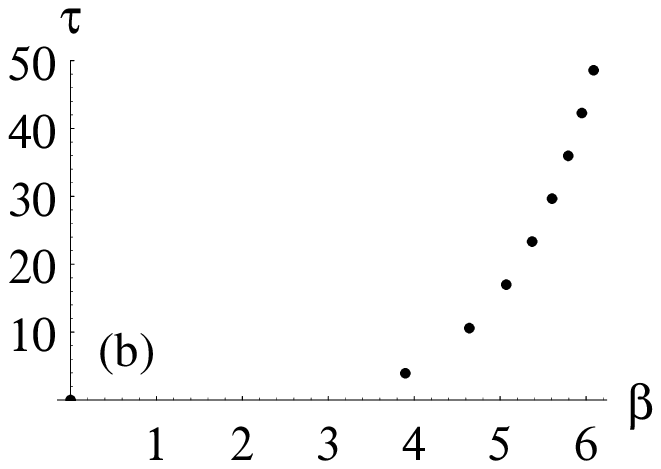 ,height=4.5 cm, angle=0}
 \caption{Complex zeros of canonical distribution function
 $\mcal{Z}_\mrm{C}(\gb +i\tau)$ for the 1D diatomic
 molecule with the piecewise linear pair interaction potential $u(r)$
 from Eq. \eqref{e:upair}. Quantities $\gb$ and $\tau$ are given in
 units of the inverse binding energy $a^{-1}$. (a) Numerically determined zeros in the upper
 half-plane obtained with Mathematica for $L/r_0=20$. (b) Analytic
 results based on the asymptotic series expansion of
 $\mcal{Z}_\mrm{C}$  at $L/r_0 \to \infty$, see Eq. \eqref{e:zeros}.  As evident from these two
 diagrams, the analytic estimate agrees well with the numerically
 determined results.  
\label{fig06}}
 \end{figure}

Figure \ref{fig05} shows several curves, corresponding to the  
thermodynamic laws~\eqref{e:EOS_canon}. One may
notice a strong local increase in the solid caloric curve of
Fig. \ref{fig05}~(a). This behavior is associated with a smooth canonical PT in the DOZ scheme of Borrmann et al. \cite{BoMuHa00}.   To see this, we next determine the complex zeros of the complex partition function  
$\mcal{Z}_\mrm{C}(\gb+i\tau,L)$ for $L>1$ from Eq. \eqref{e:Z_canon}. Figure
\ref{fig06} (a) shows the corresponding numerical results found
with Mathematica \cite{Mathematica} for $L=20$, by using a contour
plot of the function $|\mcal{Z}_\mrm{C}(\gb+i\tau,L)|$. Since the
zeros of the partition function are complex conjugate \cite{BoMuHa00},
only the zeros in the upper complex half-plane are shown.   
\par
To also obtain an analytic estimate for the DOZ in
the more interesting case $L\gg 1$, we expand the  partition function \eqref{e:Z_canon-b} near $L\to \infty$, and find
\be
\mcal{Z}_\mrm{C}(\tilde \gb,L)=\f{m\pi L^2}{h^2\tilde \gb}\left\{
e^{-\tilde \gb}\left[1 + 
\f{2}{\tilde\gb L} (e^{\tilde \gb} - \tilde\gb-1) \right]+
\mcal{O}(L^{-2})
\right\}.   
\ee
Thus, neglecting terms of $\mcal{O}(L^{-2})$, the zeros of
$\mcal{Z}_\mrm{C}(\tilde \gb,L)$ are given by
\be\label{e:zeros}
\tilde \gb_k&=&
\f{2}{L-2}-
\mrm{ProdLog\,}\biggl[-k,\f{2}{L-2}\exp\biggl(\f{2}{L-2}\biggr)\biggr],\\
k&=&\pm 1,\pm 2, \ldots,\notag
\ee
where $\mrm{ProdLog\,}[k,z]$ is the $k$th solution for $w$ in
$z=w\exp(w)$. The result \eqref{e:zeros} is valid for $L\gg 1$, and in Fig. \ref{fig06}
(b) we plotted $\tilde \gb_k$ for $1\le k\le 8$. By comparing
Figs. \ref{fig06} (a) and (b) it becomes evident that for $L\ge 20$ 
the analytic estimate from Eq. \eqref{e:zeros} is in good agreement with
the numerical results. 
\par
Using Eq.~\eqref{e:zeros} we can calculate the characteristic
quantities $\ga,\gb_c,\gc$ from Eqs.~\eqref{e:Borrmann}, required for the
DOZ classification. Table \ref{table_1} shows a summary of the results
for different values of the volume parameter $L$. The critical temperature is given by $kT_c=1/\gb_c$ and $E_c$ is obtained by inserting  $\gb_c$ into the energy equation \eqref{e:EOS_canon-a}. According to the DOZ scheme, for $1\ll L
<\infty $ we find $0<\ga<1$, corresponding to a canonical second-order PT. However, for $L\to \infty$ we observe that
$\ga\to 0$, $\gc\to 0$ indicating that in this limit the transition
converges to first order. Moreover, as evident from
the last two columns in Tab. \ref{table_1}, for increasing volume $L$ the critical temperature vanishes, $T_c\to 0$, while the
corresponding energy values $E(T_c)$ approach the value 0.5. However,
by \emph{first} taking the limit $L\to \infty$ in Eq. \eqref{e:EOS_canon-a}, we
find that the corresponding asymptotic caloric curve is given by 
\be\label{e:asymptotic}
\lim_{L\to \infty} E=1+\gb^{-1}=
1+k T, \qquad T>0.
\ee
Inserting $T_c$ into the rhs. of this equation and letting $T_c\to 0$,
we obtain $\bar{E}_c=1$, which is in agreement with microcanonical
result $E_c(L=\infty)=1$, and corresponds to the dissociation energy. 
In particular, the latter result means that, in the case of a very large volume, very small energy fluctuations (requiring $T>0$)
suffice to permanently break up the molecule. 
On the other hand, the system becomes deterministic at $T=0$, and,
correspondingly, the mean energy of the CE is then
always given by the groundstate value $E_0=0$, representing the bound
state [formally this corresponds first taking the limit $T\to 0$ in Eq.~\eqref{e:EOS_canon-a}]. 
The apparent convergence to $0.5$ in the last column of Table
\ref{table_1} just reflects the fact that for
Eq.~\eqref{e:EOS_canon-a} the two limits  $T\to 0$ and
$L\to\infty$ do not commute. 

\begin{table}[h]
\centering
\begin{tabular}{lccrcc}
\hline
\hline
$L\; [r_0]$ & $\ga$ & $\gc$ & $\gb_c\; [a^{-1}]$ & $k T_c\; [a]$ &  $E(T_c)\; [a]$ \\
\hline 
$2\cdot 10^1$  & 0.054& 0.112 & 3.5 & 0.290 & 0.922 \\
$2\cdot 10^3$  & 0.052& 0.052 & 9.0 & 0.111 & 0.688 \\
$2\cdot 10^5$  & 0.034& 0.028 & 14.1& 0.071 & 0.620 \\
\hline
$2\cdot 10^{10}$  
& $1.1\cdot 10^{-2}$ & $9.1\cdot 10^{-3}$  & 26.3   & 0.0381 & 0.56210\\
$2\cdot 10^{100}$ 
& $2.4\cdot 10^{-5}$ & $ 1.1\cdot 10^{-4}$ & 235.7 & 0.0042 & 0.50643\\
$2\cdot 10^{1000}$ 
& $2.5\cdot 10^{-8}$ & $1.2\cdot 10^{-6}$ & 2310.3 & 0.0004 & 0.50065\\
\hline
\hline
\end{tabular}
\caption{Canonical PT parameters according to the DOZ
scheme \cite{BoMuHa00} for the 1D diatomic molecule with
piecewise linear interaction potential from Eq. \eqref{e:upair}.  For
$L\to \infty$ the numerical results suggest that $\ga\to 0$, $\gc\to
0$ and $k T_c/a\to 0$.
\label{table_1}} 
\end{table}
    
\section{Summary and discussion}
\label{s:summary} 

In this paper we have studied phase  
transition-like phenomena in small systems, characterized by a finite number of degrees of freedom (DOF). The main objective was to clarify similarities and differences that arise when considering either the microcanonical ensemble (MCE), corresponding to a thermally isolated system, or the canonical ensemble (CE), corresponding to a system in contact with an infinite heat bath.

\subsection{Microscopic phase transitions in the microcanonical ensemble}

In Sec. \ref{s:pt_micro} it was shown that the
\emph{microcanonical} thermodynamic functions (TDFs) of a small system can
exhibit complex oscillatory behavior and singularities
(non-analyticities). Analogous to \emph{macroscopic} phase transitions (PTs), such non-analytic points can be 
interpreted as \emph{microscopic} PTs in the MCE.\footnote{Formally, such singularities can be softened by using an \lq artificially\rq\space smoothened box potential. However, this would \emph{not} affect experimentally observable effects, as, e.g., a drop-off of the temperature (mean kinetic energy) when the energy level for the next dissociation step is crossed. Loosely speaking, employing a smoothened box potential in the thermodynamic analysis of a small system is similar to avoiding the thermodynamic limit when being interested in PTs of large systems.} 
To illustrate the physical meaning of such microscopic PTs, we calculated the microcanonical TDFs
for two slightly different 1D toy models (Lennard-Jones and Takahashi gas). For both systems one can identify critical energy curves $E_c(L)$ along which the primary  
microcanonical thermodynamic potential -- the Hertz entropy -- is 
non-analytic, reflected by kinks or discontinuities in the TDFs.  In these models the microscopic PTs separate energetically different dissociation states. Their number and formal order increases with increasing particle number $N$ \cite{KaSc05}. In general, our results indicate that, \emph{typically, microscopic PTs in the MCE are accompanied by strong qualitative changes of the
thermodynamic observables, as e.g. rapid drop-offs or oscillations
of temperature and heat capacities}, see Fig.~\ref{fig03} (a). These effects should be
observable in suitably designed evaporation experiments, realizing the
conditions of the MCE (similar to those of Schmidt
et al. \cite{Sc01}, but without heat bath).

For the model systems considered in this paper, a non-analyticity in
the TDFs is characterized by a critical energy curve $E_c(L)$. Mathematically, such critical curves $E_c(L)$ arise due to the integration over the $\Theta$-function in
the definition of the microcanonical phase volume $\Go$. The $\Theta$-function in
Eq. \eqref{e:omega-1} effectively constrains the range of the
integration to the subset
\be\label{e:criterion} 
\mcal{A}=
\left\{
(q_1,\ldots,q_d)\in \mbb{R}^d\;|\; E-U(q_1,\ldots,q_d;V) \ge 0\right\}
\ee
in the configuration space; i.e., $\mcal{A}$ is the
energetically permitted configuration space region. 
Hence, a singularity in the microcanonical TDFs may arise
whenever $\mcal{A}$ changes its geometry or `shape' in an irregular manner during
a small variation of the control parameters $E$ and $V$. One possible
origin for this may be
a change in the topology of $\mcal{A}$ (as discussed by Pettini et
al. \cite{FrPe05a,FrPe05b,AnEtAl05}). For the  
models considered here, however, the topology of $\mcal{A}$ remains unaffected and another general mechanism is at work. To illustrate this in more detail, we recall the
example from Sec.~\ref{s:micro_example_1}, corresponding to the two-particle Lennard-Jones gas. In this case, the set $\mcal{A}$ can be expressed as
$$
\mcal{A}=
\left\{(q_1,q_2)\in \mbb{R}^2 \;|\; 
(q_1,q_2)\in \left[-L/2,L/2\right]^2 \;\wedge\; E-U_\mrm{LJ}(q_1-q_2) \ge 0\right\},\notag
$$
where $U_\mrm{LJ}$ is given by Eq. \eqref{a-e:LJ}. The first constraint for $(q_1,q_2)$ reflects the box potential, whereas the second constraint arises from the interaction potential. For $E<E_c(L)$, $\mcal{A}$ consists of the two diagonal \lq strips\rq
\be\label{eq:LJ_strips}
\mcal{A}_\pm=
\left\{
(q_1,q_2)\in \left[-L/2,L/2\right]^2
\;\wedge\;
 r_-(E)\le\pm\left(q_1-q_2\right)\le r_+(E)\right\}\;,
\ee
where $r_-(E)>0$ and $r_+(E)>0$ denote the classical turning points of the LJ potential. The \lq strips\rq\space $\mcal{A}_\pm$ are bounded by the box potential on two sides, whereas the other two boundaries are determined by the LJ interaction potential. For $E>E_c(L)$, however, the regions $\mcal{A}_\pm$ become triangles, bounded by the box potential on two sides and by the interaction potential on the remaining side:
\begin{align}\notag
\mcal{A}_\pm=
\left\{
(q_1,q_2)\in \left[-L/2,L/2\right]^2
\;\wedge\;
  r_-(E)\le\pm\left(q_1-q_2\right)
\right\}\;.\notag
\end{align}
Evidently, $\mcal{A}$ changes its geometry dramatically, when the energy passes through the critical energy $E_c(L)$, thereby giving rise to the 
microscopic PT. For a sufficiently large volume $L\gg r_0$, where $r_0$ is the range of
the pair interaction, this non-analytic transformation of $\mcal{A}$ at the dissociation energy is accompanied by a change of $\mcal A$'s effective
dimensionality. Here, we mean by `effective dimensionality'  the
number of orthogonal configuration space 
directions in which the set $\mcal{A}$  extends comparably to the
system size $L$. For energy values $E \ll E_c(L)$, the \lq strips\rq\space  $\mcal{A}_\pm$ defined by Eq.~\eqref{eq:LJ_strips} are very narrow compared to the system size $L$, and, hence, $\mcal{A}=\mcal{A}_+\cup\mcal{A}_-$ is effectively one-dimensional. When the energy
approaches $E_c(L)$ from below, the (average) width of the two subsets  $\mcal{A}_\pm$ grow very rapidly, and for $E=E_c(L)$, they become triangles whose size is of the order of $L^2$; i.e., the set $\mcal{A}$ is effectively two-dimensional for $E\ge E_c(L)$. In particular, it is this very rapid growth of $\mcal A$ -- or $\Go$, respectively --  which leads to a negative slope of the caloric curve $T(E,V)$ in the vicinity of the dissociation energy~$E_c(L)$.\footnote{This also explains why microscopic PTs appear more pronounced for larger $L$.}

\subsection{Comparison with the canonical ensemble}        
In contrast to the microcanonical TDFs, singular PTs
cannot occur in the CE of a finite system
\cite{YaLe52,LeYa52,GrRo67,GrRo69,GrLe69,Ho50,CuSa04}. This is a
consequence of the two completely different physical conditions, underlying MCE and CE, respectively
\cite{Be67,Gr01}. In spite of lacking sharp transitions, it is useful to 
distinguish different thermodynamic \lq phases\rq, when considering CEs of finite
systems \cite{BePe94,VaVlPeFo02}. Following the proposal of Borrmann  et al.  \cite{BoMuHa00}, we determined the distribution
of complex zeros (DOZ) for the canonical partition function of the two-particle  Takahashi gas (Sec. \ref{s:pt_canonical}) and found a \lq smooth\rq\space canonical second-order PT (according to the
DOZ classification scheme), provided the volume $L$ is large compared
to the range of the potential, but still finite; for $L\to \infty$ the canonical PT changes to first order in the DOZ scheme.  
\par
In principle, however, there exist important differences between the smooth PTs in the
CE and the singular microscopic PTs in
the MCE: Smooth canonical PTs 
can be viewed as direct finite size counterparts of macroscopic PTs. In
particular, for $N$ particle systems as discussed in
this paper the DOZ-scheme usually yields only one transition
point. Furthermore, the DOZ-order of the transition point is typically close to one or two. By contrast, when considering the
corresponding isolated system (i.e. the MCE), both the number and the formal order
of the singular microscopic PTs increase approximately proportional to 
the particle number (since microscopic PTs signal each dissociation
step separately). Such differences notwithstanding, canonical and microcanonical partition functions are linked by a Laplace
transformation, which suggests that (not only for small systems) there
might exist a direct connection between the canonical DOZ and the appearance
of microscopic non-analyticities in microcanonical entropy (see
also \cite{HiDu05} for a discussion of this hypothesis).  
\par 
To conclude with, if one wishes to describe small systems by means of a thermodynamic approach, then the non-equivalence of the different
statistical ensembles \cite{CoEtAl05} renders necessary to specify in
advance, whether or not a heat bath (thermostat) is coupled to the small
system under consideration. The discussion in the present paper has focussed on  two extreme limit cases, corresponding to a vanishing heat bath (MCE) and an infinite heat bath (CE). In the future, it would also be interesting to study phase transition-like phenomena in small systems coupled to a finite heat bath \cite{Ra95,ChGu02,AdEtAl03}, e.g., on the basis of Tsallis' generalized statistics \cite{Ts88,CuTs91,Gr02PhysA}.

The authors thank  W. Ebeling, D.~H.~E. Gross, P. H\"anggi, P. Talkner and an anonymous referee for numerous helpful remarks and suggestions.   

\bibliography{TD,FracCalc,Tsallis}
\bibliographystyle{elsart-num}

\end{document}